\documentclass[a4paper]{spie}  

\usepackage{amsmath,amsfonts,amssymb}
\usepackage{graphicx}
\usepackage[colorlinks=true, allcolors=blue]{hyperref}
\newcommand{\roundimg}[1]{%
  \setbox0=\hbox{\includegraphics[width=\linewidth]{#1}}%
  \begin{tikzpicture}
    \clip[rounded corners=3pt] (0,0) rectangle (\wd0,\ht0);
    \node[anchor=south west, inner sep=0pt] at (0,0) {\box0};
  \end{tikzpicture}%
}
\usepackage{tikz}
\usetikzlibrary{arrows.meta, fit, positioning, backgrounds, calc, shapes.geometric, decorations.pathreplacing}
\usepackage[T1]{fontenc}
\usepackage{listings}
\usepackage{siunitx}
\colorlet{json-string}{black!55}
\colorlet{json-number}{black}
\colorlet{json-bool}{black}

\lstdefinelanguage{json}{
  basicstyle=\ttfamily\footnotesize,
  breaklines=true,
  breakatwhitespace=false,
  columns=fullflexible,
  keepspaces=true,
  frame=single,
  framesep=4pt,
  xleftmargin=4pt,
  xrightmargin=4pt,
  showstringspaces=false,
  stringstyle=\color{json-string},
  string=[b]",
  literate=
    *{0}{{{\color{json-number}0}}}{1}
     {1}{{{\color{json-number}1}}}{1}
     {2}{{{\color{json-number}2}}}{1}
     {3}{{{\color{json-number}3}}}{1}
     {4}{{{\color{json-number}4}}}{1}
     {5}{{{\color{json-number}5}}}{1}
     {6}{{{\color{json-number}6}}}{1}
     {7}{{{\color{json-number}7}}}{1}
     {8}{{{\color{json-number}8}}}{1}
     {9}{{{\color{json-number}9}}}{1}
     {:}{{{\color{black}{:}}}}{1}
     {,}{{{\color{black}{,}}}}{1}
     {true}{{{\color{json-bool}true}}}{4}
     {false}{{{\color{json-bool}false}}}{5}
     {null}{{{\color{json-bool}null}}}{4},
}
\lstdefinestyle{jsonl}{language=json}

\usepackage{xcolor}

\title{Astra: an open-source fully autonomous robotic observatory control software}

\author[a,b]{Peter P. Pedersen}
\author[a]{David Degen}
\author[c]{Lionel Garcia}
\author[e]{Urs Schroffenegger}
\author[d]{Daniel Sebastian}
\author[c]{Sebasti\'{a}n Z\'{u}\~{n}iga-Fern\'{a}ndez}
\author[d]{Brice-Olivier Demory}
\author[g]{Elsa Ducrot}
\author[c]{Michaël Gillon}
\author[b]{Matthew J. Hooton}
\author[b]{Clàudia Janó-Muñoz}
\author[h]{James McCormac}
\author[f]{Mathilde Timmermans}
\author[f]{Amaury H.M.J. Triaud}
\author[a,b]{Didier Queloz}

\affil[a]{ETH Z\"{u}rich, Department of Physics, Wolfgang-Pauli-Strasse 2, 8093 Zurich, Switzerland}
\affil[b]{Cavendish Laboratory, JJ Thomson Avenue, Cambridge, CB3 0HE, UK}
\affil[c]{Astrobiology Research Unit, Universit\'{e} de Li\`{e}ge, All\'ee du 6 ao\^ut 19, 4000 Li\`{e}ge, Belgium}
\affil[d]{Thüringer Landessternwarte Tautenburg, Sternwarte 5, 07778 Tautenburg, Germany}
\affil[e]{Center for Space and Habitability, University of Bern, Gesellschaftsstrasse 6, 3012 Bern, Switzerland}
\affil[f]{School of Physics and Astronomy, University of Birmingham, Edgbaston, Birmingham B15 2TT, UK}
\affil[g]{AIM, CEA, CNRS, Université Paris-Saclay, Université de Paris, F-91191 Gif-sur-Yvette, France}
\affil[h]{Department of Physics, University of Warwick, Coventry, CV4 7AL, UK}

\authorinfo{Further author information: (Send correspondence to P.P.P.)\\
P.P.P.: E-mail: ppedersen@ethz.ch}

\begin{document}
\maketitle

\renewcommand{\thefootnote}{\roman{footnote}}

\begin{abstract}
Robotic and autonomous observatories are critical for modern time-domain and high-cadence astronomical surveys. The operation of these facilities requires complex software coordination to manage hardware, schedule observations, and ensure safety. However, existing observatory control software are often proprietary and platform-locked or require complex message-brokering infrastructure.
Here we present Astra (Automated Survey observaTory Robotised with Alpaca): an open-source, cross-platform Python system for the sustained, fully autonomous operation of astronomical observatories, requiring no external message-broker infrastructure. Astra controls observatory hardware via the ASCOM Alpaca protocol, and executes prescheduled observatory actions under continuous safety supervision. Its multi-device actions include plate-solve-based pointing correction with a local Gaia--2MASS catalogue fallback, PID-controlled autoguiding, and autofocus. A FastAPI web service provides a browser UI, REST and WebSocket APIs for real-time status, image previews, and SQLite-backed telemetry and logs.
Astra has run in fully unattended production since January 2024, scaling to six telescopes across three facilities: the SPECULOOS-South network (4 $\times$ 1\,m class, Chile), SAINT-EX (1\,m class, Mexico), and the ETH Observatory (0.5\,m class, Switzerland), with no schedule aborts attributable to Astra software. Across the SPECULOOS-South network, it achieves sub-arcsecond autoguiding (0.11\unit{\arcsecond} median pointing scatter) and plate-solve failure rates below 1\% on three of the four telescopes (3\% on the narrowest-field unit), demonstrating that an open, standards-based software stack can meet the reliability demands of production survey astronomy.

\end{abstract}

\keywords{observatory control software, ASCOM, Alpaca, Python, autonomous, robotic, SPECULOOS, near-infrared, bad pixels}

\section{INTRODUCTION}
\label{sec:intro}

In the field of astronomical instrumentation, an operational distinction is usually made between \emph{automated} systems, which execute rigid, predefined sequences of hardware actions; \emph{robotic} facilities, which utilise closed-loop feedback from environmental sensors or data streams to alter real-time tracking and imaging behaviours; and fully \emph{autonomous} observatories, which possess the high-level logic required for multi-night scheduling, continuous dynamic safety assessments, and independent error recovery over extended periods without human intervention.

The foundations of automated astronomy date back to pioneering facilities like the Wisconsin Automatic Photoelectric Telescope (APT) in 1965~\cite{Code_1992} and the digitally controlled Automated Supernova Search instruments developed throughout the 1970s~\cite{Colgate_1982}. The Wisconsin APT successfully executed closed-loop robotic routines, utilising automated search grids to re-centre stars on its photometer. However, it relied on operators to initiate nightly tasks. By the 1980s, the application of single-board microcomputers and integrated external weather interlocks at facilities like the Fairborn Observatory enabled true unattended robotic operations on-site~\cite{Boyd_1984,Helmer_1985}.
The scientific requirement to respond to prompt optical emission from transient gamma-ray bursts (GRBs) forced a shift in autonomous orchestration capability. The Robotic Optical Transient Search Experiment (ROTSE-I) demonstrated a major step change in software integration by ingesting streaming socket alerts directly over the GRB Coordinates Network (GCN) via TCP/IP, autonomously interrupting ongoing tasks to slew and capture contemporaneous optical radiation within seconds of a trigger~\cite{Akerlof_1999}. This operational agility paved the way for responsive distributed systems such as ROTSE-III~\cite{Akerlof_2003}, BOOTES~\cite{CastroTirado2004}, and PROMPT~\cite{Reichart_2005}.
Robotic and autonomous architectures subsequently expanded from reactive transient follow-up into systematic exoplanet transit photometry~\cite{Pollacco_2006} and wide-field synoptic surveys~\cite{Bellm_2019}. For photometric surveys in particular, full operational autonomy maximises scientific yield by instantly exploiting brief weather windows and enabling coordinated multi-site networks without a proportional, cost-prohibitive scale-up in staffing. The Liverpool Telescope~\cite{Steele_2004} and Las Cumbres Observatory (LCO)~\cite{Brown_2013} demonstrated the long-term viability of queue-scheduled robotic operations across global geographical footprints.

Building and maintaining observatory control software (OCS) is a substantial engineering undertaking, and most facilities have addressed it in one of two ways.
Larger, well-funded observatories typically develop bespoke systems tailored to their specific hardware configurations and science case. The Liverpool Telescope\cite{Steele_2004} and the Vera C. Rubin Observatory\cite{10.1117/12.2630226} are representative examples, each with control systems not designed for reuse at other facilities.
Las Cumbres Observatory (LCO)\cite{Brown_2013} occupies a hybrid position: While it demonstrated queue-scheduled robotic operations across multiple sites and open-sourced parts of its observatory management stack, including software for observation submission, scheduling, and archive-related workflows, the underlying low-level telescope and instrument control layers remain proprietary.
Conversely, smaller survey programmes have traditionally adopted commercial packages such as ACP Expert Observatory Control Software\footnote{\url{https://svn.dc3.com/}}, MaxIm DL\footnote{\url{https://diffractionlimited.com/product/maxim-dl/}}, Voyager\footnote{\url{https://software.starkeeper.it}}, and NINA\footnote{\url{https://nighttime-imaging.eu}}, adapting them to their scientific programme at hand.
These commercial solutions rely on a hardware abstraction layer to communicate with instruments in a device-agnostic way. The three dominant standards in this domain are ASCOM\footnote{\url{https://ascom-standards.org/}}, a driver framework historically built on the Windows Component Object Model (COM) and often considered the standard for amateur to entry-level professional instrumentation; INDI\footnote{\url{https://indilib.org/}}, a distributed client-server architecture favoured under POSIX operating systems (Linux, BSD, and macOS) that is widely used for embedded systems like Arduino; and INDIGO\footnote{\url{https://www.indigo-astronomy.org}}, a newer, next-generation layered architecture designed as an open-source C/C++ evolution of the INDI protocol.
To introduce native cross-platform capability, the ASCOM team introduced ASCOM Alpaca around 2018. Alpaca completely decoupled the architecture from the local operating system by replacing the COM layer with a modern, network-based HTTP/REST web-service API. This architectural shift extended universal ASCOM compatibility to Linux, macOS, and edge computing nodes, leading alternative frameworks like INDI to expand their support to include Alpaca-native hardware streams.

Beyond commercial OCS, several universal open-source solutions exist, yet they present architectural bottlenecks for modern autonomous survey operations. The foundational C++ based RTS2 framework~\cite{kubanek2004rts2} features a performant distributed architecture, but relies heavily on independently written monolithic drivers; constrained by a shrinking development community, its device support is frequently insufficient for modern, commercial-off-the-shelf (COTS) hardware models. More recently, open-source frameworks such as pyobs~\cite{Husser_2022} successfully demonstrated that a modular, Python-based OCS was achievable. However, pyobs' core implementation requires an external, separately operated XMPP messaging chat server to broker commands, and its internal ASCOM Alpaca support was tightly coupled to a specific instrument configuration rather than intended for general ecosystem reuse.

The real-world vulnerabilities imposed by this fractured software ecosystem are illustrated by the observatory control of SPECULOOS. SPECULOOS~\cite{delrez2018speculoos} (Search for habitable Planets EClipsing ULtra-cOOL Stars) is a photometric survey targeting nearby ultra-cool dwarf stars for transiting terrestrial planets and initially adopted the commercial software approach for its network of 1~m robotic telescopes. The facility utilised standard ASCOM drivers for device-level communication, but tied its top-level orchestration to ACP~\cite{Sebastian_2021}, a Windows-centric, proprietary commercial system.
Routine nightly operations relied on pre-generated observing scripts, and a custom external weather-monitoring script resumed observations after each weather interruption.
Each night, an operator connected via VPN to confirm the system state and manually launch the observing plan. As the network grew to four telescopes, this manual step consumed an increasing share of staff time. Any fault outside the scripted recovery paths aborted the sequence until an operator intervened remotely.
The vulnerability of relying on closed-source, platform-locked middleware became untenable with the announcement of ACP's end-of-support phase in 2025.
The necessity to replace this core infrastructure established a clear set of requirements for a modern, production-grade OCS: it must achieve sustained, multi-week unattended operation, recover automatically from unscripted hardware or network faults without human oversight, and enable the seamless addition of telescope nodes without a proportional increase in operational staffing. Furthermore, the framework had to be open-source and cross-platform to minimise long-term maintenance overhead, while natively supporting the ASCOM Alpaca protocol to leverage the network's existing instrument drivers and completely sever the system's dependency on the Windows operating system.

To fulfil these precise architectural demands, we developed Astra\footnote{\url{https://github.com/ppp-one/astra}} (Automated Survey observaTory Robotised with Alpaca).\cite{https://doi.org/10.5281/zenodo.18890151} Astra is an open-source, cross-platform Python-based OCS designed from the ground up for resilient, fully autonomous survey operations without the need for complex, external message-brokering infrastructure. 
It is currently deployed at SPECULOOS-South (four 1~m telescopes, ESO Paranal, Chile), SAINT-EX (San Pedro M\'{a}rtir, Mexico), and the ETH Observatory (Zurich, Switzerland).

The remainder of this paper is structured as follows. Section~\ref{sec:overview} describes the runtime architecture, including the multi-process device model, the web service, and the software stack. Section~\ref{sec:features} presents the principal features: the safety watchdog, the scheduler, plate solving, autofocus, autoguiding, deferred FITS header completion, site-specific subclassing, and the integration test suite. Section~\ref{sec:deployment} reviews production deployment experience across the three operational sites. Section~\ref{sec:performance} reports the on-sky performance of autoguiding, and plate solving. Section~\ref{sec:conclusions} summarises the work.

\section{SYSTEM OVERVIEW}
\label{sec:overview}

Astra is a cross-platform Python application that runs on Linux, macOS, and Windows. By communicating with all observatory hardware exclusively through ASCOM Alpaca HTTP/REST servers, Astra decouples its high-level orchestration logic from operating-system-specific hardware drivers. The top-level architecture of this multi-layered framework is illustrated in Figure~\ref{fig:architecture}.

\begin{figure}[t]
\centering
\resizebox{\columnwidth}{!}{%
\begin{tikzpicture}[
  every node/.style = {font=\small},
  B/.style  = {draw, thick, rounded corners=2pt, align=center,
               minimum height=0.78cm, inner sep=4pt},
  Bg/.style = {B, fill=gray!18},
  Bw/.style = {B, fill=white},
  Bs/.style = {B, fill=gray!8, font=\footnotesize},
  ar/.style = {->, >=Stealth, semithick},
  al/.style = {<-, >=Stealth, semithick},
  alr/.style = {<->, >=Stealth, semithick}]

\node[Bg, minimum width=18.0cm] (CLI)
    {Browser, REST and WebSocket clients};
\node[Bw, minimum width=18.0cm, below=0.7cm of CLI] (API)
    {\texttt{FastAPI} application};
\node[Bw, minimum width=18.0cm, below=0.6cm of API] (OBS)
    {\texttt{Observatory}};

\node[Bw, minimum width=2.5cm, below=0.9cm of OBS, xshift=-7.5cm]
    (SCH) {\texttt{ScheduleManager}};
\node[Bw, minimum width=2.5cm, below=0.9cm of OBS, xshift=-4.5cm]
    (THM) {\texttt{ThreadManager}};
\node[Bw, minimum width=2.5cm, below=0.9cm of OBS, xshift=-1.5cm]
    (GDM) {\texttt{GuiderManager}};
\node[Bw, minimum width=2.5cm, below=0.9cm of OBS, xshift=+1.5cm]
    (SAF) {\texttt{SafetyMonitor}};
\node[Bw, minimum width=2.5cm, below=0.9cm of OBS, xshift=+4.5cm]
    (QUE) {\texttt{QueueManager}};
\node[Bw, minimum width=2.5cm, below=0.9cm of OBS, xshift=+7.5cm]
    (DVM) {\texttt{DeviceManager}};

\node[Bw, minimum width=2.5cm, below=0.8cm of QUE]
    (DBM) {\texttt{DatabaseManager}};

\node[Bg, minimum width=2.5cm, below=0.7cm of DBM] (DB) {SQLite};
\node[Bg, minimum width=2.5cm] (FITS) at (SCH |- DB) {FITS files};

\node[Bs, minimum width=2.5cm, minimum height=1.2cm,
      below=3.9cm of DVM, align=center]
    (SUB) {Device subprocesses\\
           Camera \quad Telescope\\
           Focuser \quad Dome \quad \ldots};

\node[Bg, minimum width=18.0cm, below=2.8cm of DB, xshift=-4.4cm] (ALP)
    {ASCOM Alpaca HTTP/REST device servers};

\begin{scope}[on background layer]
  \node[draw=gray!55, dashed, rounded corners=6pt, inner sep=10pt,
        fit={(API)(OBS)(SAF)(SCH)(THM)(QUE)(DVM)(GDM)(DBM)(DB)(FITS)}]
        (MP) {};
\end{scope}
\node[font=\scriptsize\itshape, gray!70, fill=white, inner sep=1pt]
    at ($(MP.north east)+(-0.95,0)$) {main process};

\draw[alr] (CLI) -- (API);
\draw[alr] (API) -- (OBS);

\coordinate (J) at ($(OBS.south)+(0,-0.30)$);
\draw[al] (OBS.south) -- (J);
\draw[ar] (J) -| (SCH.north);
\draw[ar] (J) -| (THM.north);
\draw[ar] (J) -| (SAF.north);
\draw[ar] (J) -| (QUE.north);
\draw[ar] (J) -| (DVM.north);
\draw[ar] (J) -| (GDM.north);

\draw[ar] (QUE) -- (DBM);
\draw[alr] (DBM) -- (DB);

\draw[ar, dashed, gray!60] ($(DBM.west)+(0,+0.1)$) to[out=180, in=270] (SAF.south);
\draw[alr, dashed, gray!60] ($(DBM.west)+(0,-0.1)$) to[out=180, in=270] (GDM.south);

\draw[ar] ($(OBS.south)+(-5.8,0)$)
    to[out=-90, in=90, looseness=0.75]
    node[pos=0.85, left=3pt, font=\scriptsize\itshape, align=center]
        {save (\texttt{ImageHandler},\\per camera)}
    ($(FITS.north)+(1.1,0)$);

\draw[alr] (DB.west) -- (FITS.east)
    node[midway, above=1pt, font=\scriptsize\itshape] {complete (\texttt{HeaderManager})};

\draw[alr] (DVM.south) -- (SUB.north)
    node[midway, right=-0pt, font=\scriptsize\itshape] {Pipe};
\draw[ar] ($(SUB.north)+(-1.1,0)$) to[bend right=20, looseness=0.3]
    node[pos=0.5, above right=-2pt, font=\scriptsize\itshape] {Queue}
    (QUE.east);
\draw[alr] (SUB.south) -- (SUB.south |- ALP.north)
    node[midway, right=2pt, font=\scriptsize\itshape] {HTTP};

\end{tikzpicture}}
\caption{Astra system architecture. The dashed boundary marks the main
  process; each Alpaca device runs in a separate subprocess outside it.
  The \texttt{Observatory} holds seven managers: \texttt{SafetyMonitor},
  \texttt{ScheduleManager}, \texttt{ThreadManager} (lifecycle of every
  main-process thread), \texttt{QueueManager}, \texttt{DeviceManager},
  \texttt{GuiderManager} (one \texttt{Guider} per telescope), and
  \texttt{DatabaseManager}.
  Subprocesses push polled telemetry and log records back through a
  shared \texttt{multiprocessing.Queue} owned by the
  \texttt{QueueManager}, which forwards them to the
  \texttt{DatabaseManager} for serialisation into the central SQLite
  store. A hung or faulting device therefore cannot block the main
  process or other devices. The dashed arc shows the
  \texttt{SafetyMonitor} reading \texttt{ObservingConditions} history
  through the \texttt{DatabaseManager} to evaluate internal weather
  limits. During each exposure the \texttt{Observatory} writes FITS
  files via \texttt{ImageHandler} (one per camera); the
  \texttt{complete\_headers} action then reads deferred telemetry from
  SQLite and rewrites those FITS files via \texttt{HeaderManager}.}
\label{fig:architecture}
\end{figure}

At runtime, a single \texttt{Observatory} instance manages all devices at a site. Each Alpaca device runs in a dedicated subprocess that subclasses \texttt{multiprocessing.Process} and communicates with the main process via \texttt{multiprocessing.Pipe}; the supported ASCOM device types are \texttt{Telescope}, \texttt{Camera}, \texttt{FilterWheel}, \texttt{Focuser}, \texttt{Dome}, \texttt{SafetyMonitor}, \texttt{ObservingConditions}, \texttt{CoverCalibrator}, \texttt{Switch}, and \texttt{Rotator}. This structural isolation prevents an unresponsive or faulting device from blocking the main process. Each device subprocess runs its own polling loop in an internal thread, acquiring the properties needed for FITS headers at intervals defined per device in the configuration (5\,s by default, 1\,s for safety monitors). Polled values, log records, and database writes flow back to the main process through a shared \texttt{multiprocessing.Queue}.

Two threads in the main process coordinate scheduling and execution of observations. The \texttt{ScheduleManager} watches the schedule file on disk, holds the parsed \texttt{Schedule} object, and exposes the iteration interface used by the schedule runner (Section~\ref{sec:scheduling}). The \texttt{ThreadManager} manages every other main-process thread, including the watchdog loop, the schedule runner, and one thread per running schedule action; it tracks threads so that dead threads can be detected and removed without restarting the process.

Astra is backed by a single SQLite database. The \texttt{QueueManager} thread drains the shared queue and forwards every record to the \texttt{DatabaseManager}, which serialises writes through a dedicated \texttt{Sqlite3Worker} thread. A single writer avoids the file-level locking that SQLite would otherwise impose under concurrent writes, and lets the device subprocesses, the watchdog, and the web service enqueue writes without each holding its own database connection. The FastAPI service reads from the same file through short-lived connections; SQLite runs in WAL mode, so these reads proceed concurrently with the writer.

\subsection{Web Interface}
\label{sec:webui}

The web layer is a FastAPI application served by Uvicorn. Its REST API groups endpoints into four categories: manual observatory control (toggle the robotic switch, close the dome, cool the cameras, force header completion); schedule management (read, edit, and upload the active JSONL schedule); telemetry retrieval (device polling history, guiding residuals, image lists); and real-time status (\texttt{/api/heartbeat}). Two WebSocket endpoints push real-time device status (\texttt{/ws}) and log records (\texttt{/ws/log}) to connected clients without polling. External scripts and monitoring systems can consume these endpoints directly; the browser UI is built on top of the same API.

\begin{figure*}[ht]
\centering
\begin{minipage}[t]{0.24\textwidth}
  \centering
  \roundimg{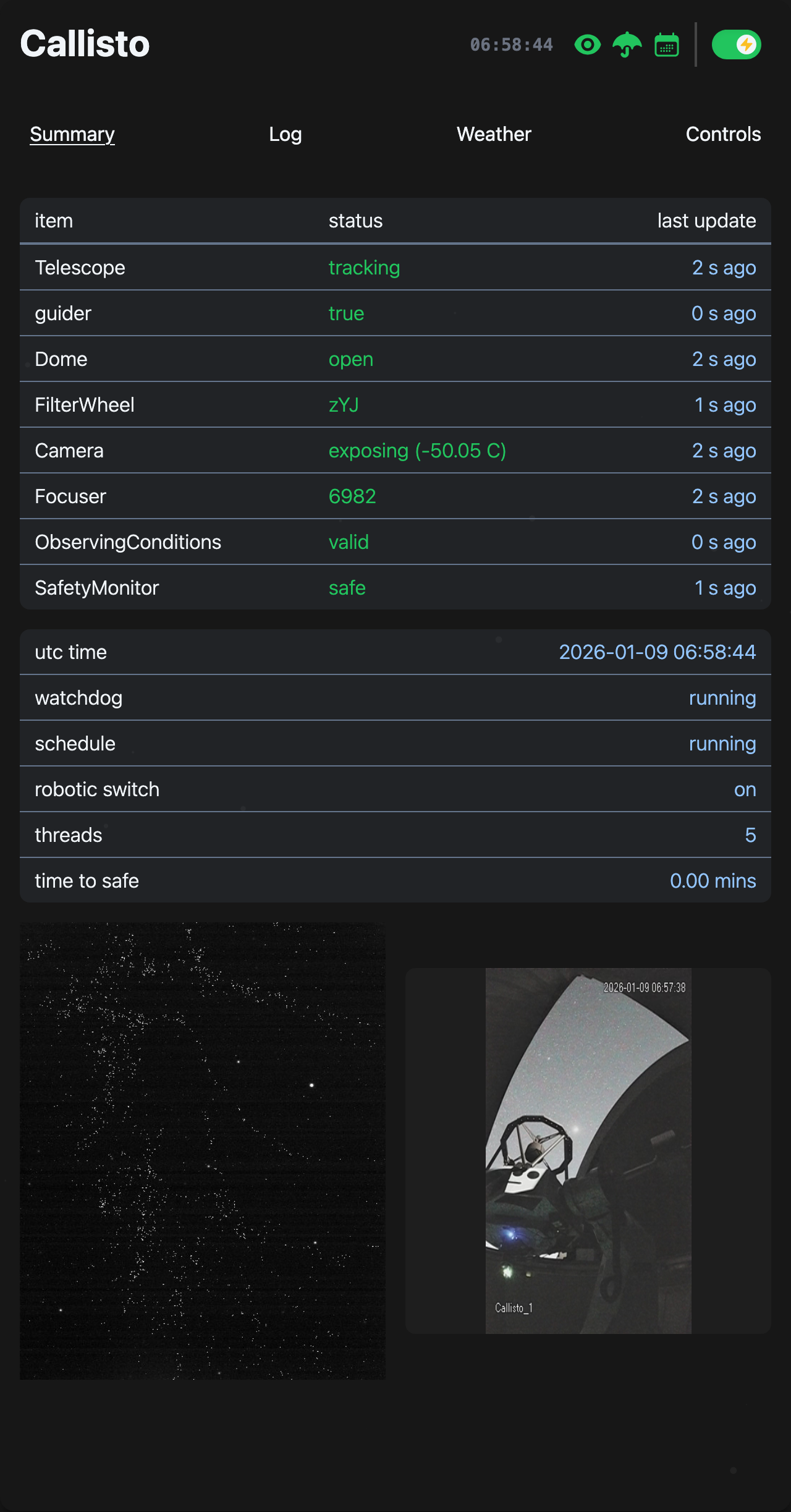}
  \small Summary
\end{minipage}\hfill
\begin{minipage}[t]{0.24\textwidth}
  \centering
  \roundimg{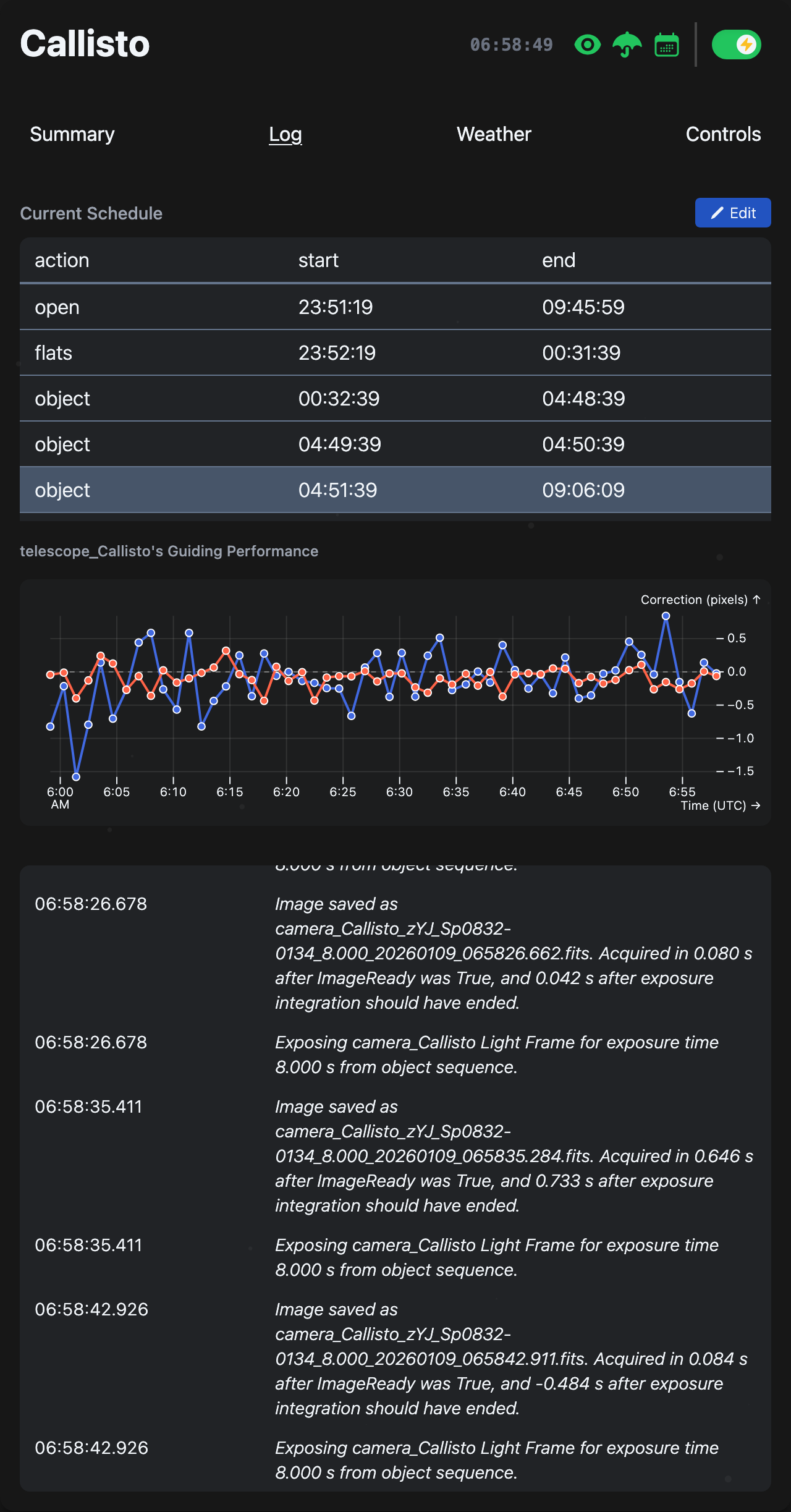}
  \small Log
\end{minipage}\hfill
\begin{minipage}[t]{0.24\textwidth}
  \centering
  \roundimg{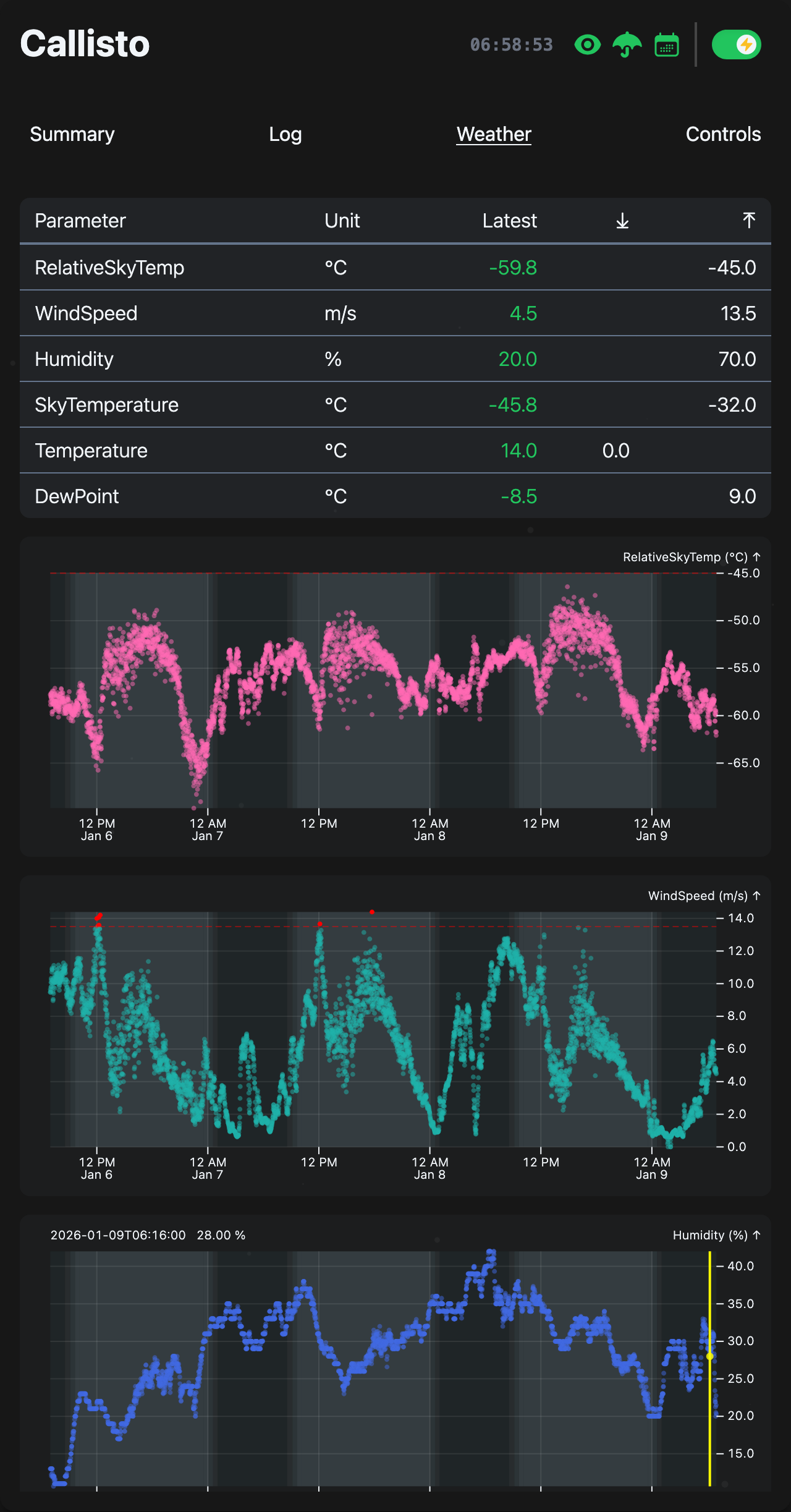}
  \small Weather
\end{minipage}\hfill
\begin{minipage}[t]{0.24\textwidth}
  \centering
  \roundimg{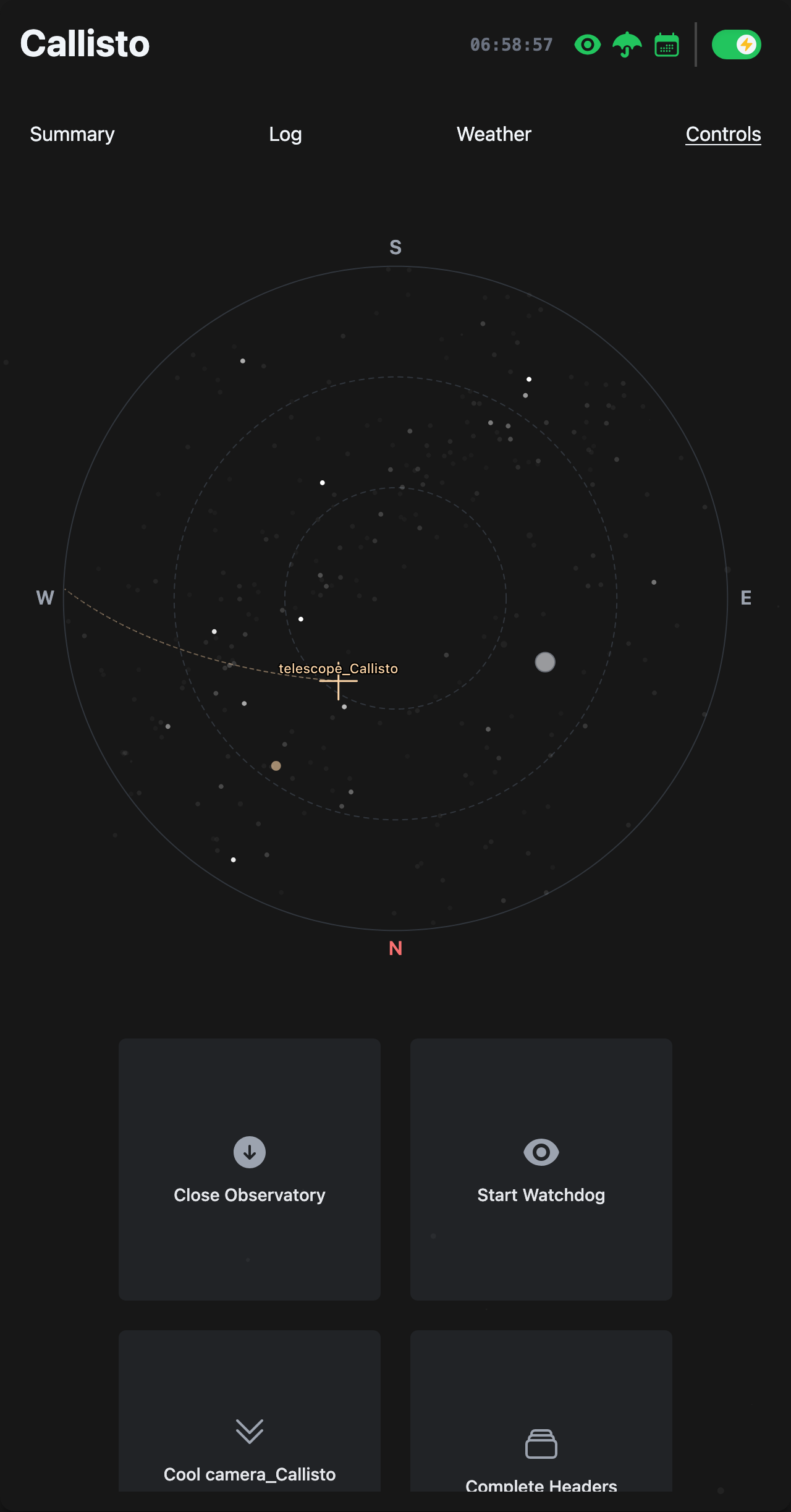}
  \small Controls
\end{minipage} \vspace{6pt} %
\caption{Astra browser interface during an active observing sequence. The Summary panel shows live device states and the most recent science frame. The Log panel streams watchdog and scheduler events in real time. The Weather panel plots time-series histories of all polled observing-conditions parameters. The Controls panel exposes manual commands for closing the observatory, cooling cameras, and triggering header completion. The same data are accessible programmatically via the REST and WebSocket APIs.}
\label{fig:webui}
\end{figure*}

The browser UI is rendered through Jinja2 templates and presents four panels. The \emph{Summary} panel displays a live device status table and the most recently acquired FITS image as a scaled JPEG. The \emph{Log} panel shows the current schedule state alongside a real-time scrolling log stream and a guiding-residual chart. The \emph{Weather} panel plots time-series histories of all polled \texttt{ObservingConditions} parameters. The \emph{Controls} panel exposes manual commands for closing the observatory, cooling cameras, and forcing header completion. It also includes an integrated schedule editor that validates each line of the active JSONL programme against the loaded observatory configuration before writing it back to disk. A dedicated \emph{FITS explorer} at \texttt{/fits\_explorer} provides in-browser image inspection with header display; all JPEG previews are generated in memory and streamed over HTTP, avoiding intermediate disk I/O. Figure~\ref{fig:webui} shows the browser interface during an active observing sequence.

\subsection{Software Stack}
\label{sec:stack}

Astra targets Python 3.11 and is distributed as a single PyPI-installable package. Its main dependencies are Astropy\cite{astropy_2022} for coordinate transforms, time-standard conversions, and FITS I/O; photutils\cite{photutils} for point-source detection; twirl\cite{twirl} for world coordinate system (WCS) computation during plate solving; \texttt{donuts}\cite{donuts} for image-to-image shift estimation in the autoguider; and \texttt{cabaret} for both online Gaia catalogue queries and the synthetic image generation used by the test simulator. The autofocus routine is provided by the companion package \texttt{astrafocus}.

Site configuration is split across two files. A global \texttt{astra\_config.yml} (stored in \texttt{\textasciitilde/.astra/}) records two paths and a name: the assets directory, the local Gaia--2MASS catalogue, and the observatory name. An observatory-level \texttt{observatory\_config.yml} in the assets directory declares every attached device under its ASCOM type key, with its Alpaca IP address, device number, and type-specific parameters. The \texttt{ObservingConditions} section also hosts the per-parameter closing limits and \texttt{max\_safe\_duration} values consumed by the safety watchdog. A companion CSV file maps each FITS keyword to its source device property, allowing sites to extend or override the header set without modifying core code. Template versions of both files are installed automatically on first run.

\section{KEY FEATURES}
\label{sec:features}

\subsection{Safety Watchdog}
\label{sec:watchdog}

Autonomous observatory operation requires a fail-safe execution loop capable of halting observations, closing the protective enclosure, and resuming operations entirely without human intervention. To ensure high hardware availability without risking catastrophic equipment damage, Astra's \texttt{SafetyMonitor} class implements an in-depth safety architecture. The primary protection layer leverages an external ASCOM \texttt{SafetyMonitor} device polled continuously at 1\,s intervals. Rather than relying on instantaneous state transitions, the system queries the monitor's \texttt{IsSafe} history over a user-defined \texttt{max\_safe\_duration} (default 15-minute) temporal window. This rolling evaluation window guarantees that a single anomalous, transient "safe" data packet cannot trigger a premature reopening cycle immediately following an adverse weather event.

While this external monitor provides an excellent global hardware-level fail-safe, a robust autonomous system requires finer operational control. Therefore, the second layer consists of a configurable internal evaluation of individual \texttt{ObservingConditions} parameters against per-parameter closing limits defined in the configuration. Each limit specifies an upper bound, a lower bound, and a \texttt{max\_safe\_duration} window over which the limit must hold; the parameters typically include wind speed, humidity, sky brightness, and the derived relative sky temperature introduced below. The \texttt{check\_internal\_conditions} method returns a safety flag and the time remaining until every limit's window is again clear.

The safety evaluation also uses a derived quantity, the \emph{relative sky temperature} $T_\mathrm{rel} = T_\mathrm{sky} - T_\mathrm{amb}$, computed at evaluation time from the polled \texttt{SkyTemperature} and \texttt{Temperature} values. Cloud detection by infrared thermopile is conventionally expressed as this difference. The absolute sky brightness in the 8--14\,\textmu m window depends strongly on ambient temperature and humidity, the relative reading gives a more stable indicator of cloud cover than the absolute value. Clear conditions correspond to a large negative offset, and overcast conditions to an offset approaching zero. 

The execution of these safety checks is managed by a dedicated, watchdog loop running inside a main-process thread. On each execution cycle, the watchdog first queries the \texttt{ScheduleManager} to determine if a local schedule file reload is required (Section~\ref{sec:scheduling}), then invokes the core \texttt{SafetyMonitor.update\_status()} routine. A \texttt{False} safety return triggers an immediate execution of the \texttt{close\_observatory()} routine. The watchdog itself can never command a reopen. Dome opening is handled inside the schedule actions: the first dome-requiring action after \texttt{update\_status()} returns \texttt{True} (after the maximum triggered \texttt{max\_safe\_duration} satisfied) calls \texttt{open\_observatory()} before slewing. Non-dome actions such as camera cooling, dark frames, and header completion can execute while the dome remains closed.

Device-level faults are detected independently of the weather-monitoring loop described above. 
When a device subprocess raises an exception, \texttt{logger.report\_device\_issue()} sets the global \texttt{error\_free} flag to false and appends the device's error metadata to the \texttt{error\_source} list. This state mutation immediately diverts the watchdog from its nominal execution cycle into \texttt{\_handle\_watchdog\_errors()}.
The watchdog then decides whether to close the dome based on which device type faulted, since closing is not always itself a safe action. Under baseline precautions, if neither the Dome nor the Telescope appears in the fault list, the observatory is closed unconditionally, treating a fault in any other device (camera, filter wheel, focuser) as grounds for a full shutdown. Conversely, if the Telescope is faulting and the Dome is not, the dome is closed only when the per-dome \texttt{close\_dome\_on\_telescope\_error} flag is enabled, protecting roll-off-roof enclosures where a moving roof could collide with a telescope that failed to park. Finally, if the Dome itself is among the faulting devices, no automatic closing is attempted; the fault is logged and left for operator intervention, since the watchdog cannot independently confirm the dome will respond to a close command it has already failed to execute correctly.

External remote monitoring applications can track this combined hardware and environmental telemetry in real time via a public \texttt{/api/heartbeat} REST endpoint.
In production setups, such as the SPECULOOS-South array deployment, a detached, external cron service queries this endpoint at regular 10-minute intervals and automatically alerts the supporting team via email if the \texttt{error\_free} state drops. Although native device-level driver faults remain exceptionally rare during standard operations (Section~\ref{sec:deployment}), this decoupled external architecture acts as an essential, high-level operational fallback loop that bridges the gap between software-level exception handling and automated human escalation.

\subsection{Scheduling}
\label{sec:scheduling}

Autonomous survey operations require a scheduling logic capable of translating high-level astronomical targets into deterministic hardware commands while handling real-time environmental interruptions. To achieve this resilience without complex middleware, Astra implements a camera-centric scheduling philosophy. At action time, the designated camera identifier is resolved against the global observatory configuration to a unified \texttt{PairedDevices} record. This record maps co-dependent hardware, such as, the telescope mount, filter wheel, focuser, and dome enclosure—directly to that specific imaging chain. The core schedule runner reads this structural pairing to dispatch concurrent Alpaca commands across network devices and dynamically construct metadata for subsequent FITS header assembly.

Observation sequences are defined within JSONL files in which each JSON line encodes a single \texttt{Action}. The supported action types are \texttt{object}, \texttt{calibration}, \texttt{flats}, \texttt{calibrate\_guiding}, \texttt{autofocus}, \texttt{pointing\_model}, \texttt{open}, \texttt{close}, \texttt{cool\_camera}, and \texttt{complete\_headers} (the deferred-header pass described in Section~\ref{sec:headers}). Camera cooling is triggered implicitly at the start of any camera-bound action; the \texttt{cool\_camera} action allows it to be scheduled explicitly when thermal stability must be established before observations begin. Each \texttt{Action} includes a camera device name (as defined in the user's observatory configuration file), the action type, a typed configuration (\texttt{action\_value}), and UTC start and end times. The following excerpt illustrates the format for an \texttt{object} action type:

\begin{lstlisting}[style=jsonl]
{
  "device_name": "camera_ir",
  "action_type": "object",
  "action_value": {
    "object":   "TOI-700",
    "ra":       97.096,
    "dec":     -65.579,
    "filter":   "H",
    "exptime":  6.00,
    "guiding":  true,
    "pointing": true
  },
  "start_time": "2025-11-29 02:00:00.000",
  "end_time":   "2025-11-29 08:14:00.000"
}
\end{lstlisting} 

Schedule validation checks filter wheel filter names against paired devices and, given the observatory's \texttt{EarthLocation}, performs altitude visibility checks for each target against a configurable minimum elevation. Invalid actions trigger alerts and are flagged as un-executable, ensuring that physical collisions or unobservable targets are filtered out before night operations begin.

JSONL was chosen over a single JSON array primarily for operational robustness and tooling convenience: each line is an independently parse-able object, so the file can be edited by hand, diffed, or appended to without rewriting a wrapping array, and a malformed or commented-out line (lines beginning with \texttt{//} are skipped) does not require touching the rest of the file. \texttt{Schedule.from\_file} reads the file in full and decodes it into a list of actions before constructing a \texttt{Schedule}. The \texttt{pointing} and \texttt{guiding} flags enable optional refinement: if \texttt{pointing=true}, the first exposure is plate-solved and, if the measured offset exceeds tolerance (user defined, 0.1' default), the telescope is slewed and the next exposure is solved again, for up to three attempts before the sequence proceeds regardless of outcome; if \texttt{guiding=true}, the autoguiding thread is started only once pointing has converged (or has been abandoned after the third attempt), and runs for the remainder of the sequence. Setting either flag to \texttt{false} skips the corresponding step entirely.

The schedule runner (\texttt{Observatory.run\_schedule}) starts a new thread for any action whose time window has opened, whose preconditions (\texttt{check\_conditions}, Section~\ref{sec:watchdog}) are met, and whose status is \texttt{pending}. By default an action's thread is joined immediately, so the next loop iteration effectively waits for it to finish before considering further actions for that schedule item. Setting \texttt{execute\_parallel: true} on an action skips this join, allowing a subsequent action (e.g. imaging on a second camera) to start while the first is still executing. 
Action failure handling is conservative: any unhandled exception raised while executing an action (e.g. a device timeout) is caught in \texttt{run\_action}, logged via \texttt{report\_device\_issue}, and sets \texttt{schedule\_manager.running} to \texttt{False}, halting the entire schedule. Recovery from error requires an operator re-enabling the watchdog and the robotic switch.

The \texttt{ScheduleManager} tracks the modification time of the schedule file and detects any change. Reload behaviour depends on whether a sequence is running. If a schedule is running when the file changes, the new file is not loaded until the running sequence finishes against its own snapshot; the watchdog then picks up the new schedule on the next loop iteration. If no schedule is running, the watchdog reloads immediately and re-starts execution under the robotic switch. Schedules can be uploaded, created, or edited through the browser UI or via the REST API.

\subsection{Plate Solving with Local Catalogue Fallback}
\label{sec:platesolver}

Reliable autonomous pointing correction must continue to work when an observatory site loses its internet connection. Astra's \texttt{pointer} module first queries reference catalogues (Gaia or 2MASS) via the online Gaia archive through the \texttt{cabaret} library, optimising for the filter present, then falls back to a local Gaia--2MASS catalogue\cite{gaia_2mass_catalogue} when the query fails or when local operation is requested explicitly. Both paths apply proper motion corrections from the Gaia reference epoch to the observation date.

Each frame is cleaned before star detection, to remove background structure and hot pixels. A 32\,$\times$\,32 background mesh is fitted with sigma-clipped median statistics and subtracted, and the residual is run through a 5\,$\times$\,5 median filter. The \texttt{DAOStarFinder} algorithm\cite{stetson_1987} from \texttt{photutils}\cite{photutils} then detects stars at a 7\,$\sigma$ threshold, with FWHM derived from the plate scale assuming 2\unit{\arcsecond} seeing. DAOStarFinder is invoked with \texttt{threshold = 7\,$\sigma_\mathrm{bkg}$}, which sets the detection significance relative to the background noise. The returned catalogue is then filtered against \texttt{peak > $\mu_\mathrm{bkg}$ + 7\,$\sigma_\mathrm{bkg}$}, which rejects any source whose central pixel does not rise above the same significance. The second cut removes spurious detections that pass the DAOStarFinder kernel, such as extended residuals from imperfect background subtraction. 

Star detection stops at the 20 brightest sources to bound the cost of the subsequent point-set matching by \texttt{twirl}, whose runtime grows steeply with catalogue size.  The \texttt{twirl}\cite{twirl} library computes the WCS transformation between pixel and Gaia queried coordinates from these stars. The solver accepts a result when 70\% of detected stars match catalogue entries within a 20-pixel threshold, or when at least eight stars match, whichever condition is satisfied first. The resulting \texttt{PointingCorrection} object stores both the target coordinates and the measured field centre, and exposes \texttt{proxy\_ra} and \texttt{proxy\_dec}: the coordinates to which the mount should slew to centre the intended target, compensating for the measured systematic offset.

\subsection{Autofocus}
\label{sec:autofocus}

To maintain optimal point-spread function (PSF) quality across varying thermal and gravitational environments, Astra implements an automated execution loop built upon the companion package \texttt{astrafocus}\cite{Degen_AstraFocus}, which provides a flexible focus-curve analysis engine and a unified registry of focus-measure operators.

Field selection is managed via \texttt{astrafocus}'s \texttt{ZenithNeighbourhoodQuery} that interfaces with the same local Gaia--2MASS catalogue that backs the plate solver. The engine queries for candidates within a configurable maximum zenith angle and a configurable Gaia G or 2MASS J magnitude range. Results are sorted and dispatched to one of three configuration-driven selection algorithms: \texttt{SINGLE} isolates an individual star with an empty neighbourhood within the field of view; \texttt{MAXIMAL} targets the pointing coordinate containing the highest absolute star density; and \texttt{ANY} accepts the first candidate passing the threshold filters. If database unavailability or a lack of suitable catalogue stars triggers a selection exception, Astra executes a robust fail-safe routine: the telescope points directly to the local zenith and continues execution using whatever field stars happen to occupy the frame.

Once a field is acquired, a focus sweep is executed. The sweep proceeds in one or more passes (configured via \texttt{n\_steps}, e.g.\ \texttt{[30, 20]}): for each pass, the focuser visits a set of evenly spaced positions across the current search range, alternating direction between passes to avoid backlash bias, and captures \texttt{n\_exposures} images at each position. Each image is reduced to a single focus-measure value; a failed measurement is recorded as missing rather than aborting the sweep. Between passes, the search range can be narrowed (\texttt{decrease\_search\_range}) before the next pass begins.

The focus per image is measured using focus measure operators from the \texttt{FocusMeasureOperatorRegistry}. Astra supports half-flux radius (HFR), two-dimensional Gaussian profile fitting, Fast Fourier Transform (FFT) sharpness metrics, normalised frame variance, and a dedicated star-size operator. While the primary operator drives the active control loop, secondary operators are computed concurrently to provide diagnostic comparison profiles. Structurally, macro-level frame-sharpness operators (FFT and normalised variance) are preferred when navigating wide search ranges where stars are heavily blurred, as they tolerate severe defocusing, whereas localised star-shape operators (HFR and Gaussian fitting) are better suited to fine-tuning near the focus peak, where point-like stellar profiles enable reliable centroiding.

The choice of focus-measure operator also determines which of two extremum-finding strategies \texttt{astrafocus} uses, selected automatically rather than configured independently. If the operator is a star-shape (\texttt{StarSizeFocusMeasure}) operator, such as HFR or the Gaussian fit, Astra instantiates an \texttt{AnalyticResponseAutofocuser}, which fits a parametric model to the accumulated focus-measure curve, solves for the extremum analytically, and narrows the search range between passes by discarding the worst-performing \texttt{percent\_to\_cut}\% of the fitted curve. For all other operators (FFT, normalised variance), a \texttt{NonParametricResponseAutofocuser} is used instead: it estimates the extremum without assuming a functional form, via one of several smoothing methods (LOWESS, median filter, spline, or RBF) set through \texttt{extremum\_estimator}; for this operator family, \texttt{decrease\_search\_range} has no narrowing effect between passes, so successive passes resample the same range at different step densities rather than zooming in.

A separate \texttt{Defocuser} class (\texttt{astra.autofocus}) supports defocused photometry of bright targets: it moves the focuser to a position offset from the stored best-focus value (\texttt{focus\_shift} or an absolute \texttt{focus\_position}) to spread a star's flux over more pixels and avoid detector saturation.

\subsection{Autoguiding}
\label{sec:guiding}

Astra defaults to autoguiding directly from the science camera frames, removing the need for a dedicated guide camera. For instrumentation configurations that explicitly require an isolated guiding channel, the architecture allows a secondary \texttt{Camera} instance to pair with the same telescope axis and execute a parallel guiding action. The guiding framework uses the \texttt{donuts}\cite{donuts} image-registration library to compute an ($x$, $y$) pixel-shift vector by cross-correlating each incoming frame against a static reference image. The resulting pixel displacement is immediately translated into a standard ASCOM pulse-guide command and dispatched to the mount.

Before passing a frame to \texttt{donuts}, the guider applies the same image cleaning methods as used by the plate solver. To strictly bound computational registration overheads, frames with dimensions exceeding 2048\,pixels are automatically centre-cropped to a 2048\,$\times$\,2048 pixel matrix, while a median filter suppresses isolated hot pixels that could otherwise bias the cross-correlation peak. 
Reference images are managed dynamically within a dedicated SQLite database table using a unique lookup key composed of a \texttt{(field, filter, exptime, camera, pier\_side)} tuple. 
Storing references per pier side provides native support for German Equatorial Mount meridian flips: Following an automated flip sequence, the guider retrieves the historical reference image recorded for the opposite pier orientation rather than forcing a fresh reference acquisition. This localized caching mechanism is actively utilized by the GEM architecture at the ETH Observatory, whereas the mounts deployed at SPECULOOS-South and SAINT-EX are configured for continuous tracking and bypass meridian flip logic entirely.

Calculated pixel shifts are converted into physical drive durations through two independent PID controllers dedicated to each orthogonal camera axis. Empirical scale mapping is established via a discrete \texttt{GuidingCalibrator} routine that calculates the pixel-to-pulse-time (\texttt{PIX2TIME}) conversion factor. This calibrator sequentially pulses the mount along the cardinal axes north, south, east, and west for a fixed duration, computes the resulting pixel displacement using an immediately preceding frame, and averages the results over a configurable number of cycles. This calibration matrix simultaneously determines the orientation mapping of the camera sensor relative to the right ascension axis (\texttt{RA\_AXIS}) and confirms the sign convention of the directional drive commands.

To mitigate high-frequency seeing variations and prevent the guider from over-correcting, consecutive adjustments are gated by a minimum temporal threshold defined by \texttt{max(MIN\_GUIDE\_INTERVAL, EXPTIME)}, where \texttt{MIN\_GUIDE\_INTERVAL} defaults to 60\,s. For short-exposure cadences, this floor suppresses rapid, closely spaced mount pulses; for long science exposures exceeding 60\,s, this pacing constraint scales naturally, yielding exactly one correction vector per completed frame. The system maintains a twenty-element rolling buffer of historical corrections per axis. Once this array is populated, the engine discards any incoming correction vector whose magnitude exceeds ten times the buffer's moving standard deviation, though the outlier is still appended to the rolling window to update baseline statistics. A strict twenty-pixel absolute command ceiling prevents corrupted frames from issuing large, destabilizing slews. This saturation cap automatically doubles to forty pixels during a three-image stabilization window following a fresh guider launch to accommodate initial settling transients.

\subsection{Deferred FITS Header Completion}
\label{sec:headers}

Writing a complete FITS header at exposure time usually demands metadata from every attached device immediately before each exposure. At sites with multiple instruments sharing an Alpaca server, these queries introduce latency between consecutive exposures and reduce on-sky efficiency. Astra addresses this by separating header data into two categories written at different times.

A per-site FITS configuration file maps each header keyword to its source. Every row is marked either \emph{fixed} or \emph{polled}: fixed values are captured at the start of each action sequence, and polled values are added to the headers at the end of a complete schedule or whenever \texttt{complete\_headers} action appears in the schedule. Astra writes a \emph{base header} at file write using values acquired at the beginning of an action sequence. It contains available fields such as target coordinates \texttt{RA} and \texttt{DEC}, set exposure duration \texttt{EXPTIME}, the standard \texttt{IMAGETYP} field, the observatory location, and timestamps derived from camera readout, together with the device properties marked fixed in the configuration. The fixed set is restricted to quantities that do not change during a sequence, e.g. the active filter name, the pixel size, the aperture diameter, the focal length, and similar instrument metadata. Properties that drift during the night, such as ambient weather, are excluded from the fixed set and populated from the polling stream instead. Each device subprocess polls its device continuously and forwards measurements through the queue to be written to SQLite with a UTC timestamp.

The schedule runner starts a \texttt{final\_headers} thread automatically at the end of every schedule run, whether or not a \texttt{complete\_headers} action appears in the schedule. \texttt{HeaderManager.final\_headers} retrieves every header-incomplete image from the database, queries the polled telemetry covering the corresponding time window, time-interpolates each polled property to the exposure start time recorded in \texttt{DATE-OBS}, updates the on-disk header, and marks the image as header complete.

\subsection{Site-Specific Customisation}
\label{sec:customisation}

The \texttt{Observatory} class is the single control point for all site-specific behaviour. To modify default behaviour, operators subclass \texttt{Observatory} within a standard Python script placed inside the user's local \texttt{custom\_observatories/} directory. At system initialisation, Astra's \texttt{ObservatoryLoader} scans that directory and selects the subclass whose class name, or whose \texttt{OBSERVATORY\_ALIASES} list attribute, matches the name supplied via the \texttt{-\--observatory} command-line flag. If no match is found, Astra uses the base \texttt{Observatory} class.

Sites most commonly replace the dome-sequence methods \texttt{open\_\-obser\-vatory()} and \texttt{close\_\-obser\-vatory()}. The SPECULOOS-South subclass is the representative example. Its \texttt{open\_\-obser\-vatory()} override calls a subclass method, \texttt{speculoos\_check\_and\_ack\_error()}, that dispatches the Astelco-specific \texttt{CommandString} calls needed to clear latched AsTelOS error states before motion is re-enabled, then blocks until the mount reports readiness via the base-class \texttt{\_wait\_for\_telescopes\_ready()} helper. Isolating these proprietary vendor instructions entirely within the local subclass keeps the core framework repository agnostic and globally reusable, enabling the SPECULOOS deployment to automatically recover from critical firmware-level faults that previously required manual human intervention.

\subsection{Testing and Simulation}
\label{sec:testing}

To verify the architectural stability of Astra, it incorporates an integration test suite that exercises the entire control stack against an isolated, end-to-end simulated observatory. The backend simulation layer relies on the \texttt{alpaca-simulators} companion package\footnote{\url{https://github.com/ppp-one/alpaca-simulators}}, which executes as an independent ASCOM Alpaca HTTP server bound to \texttt{localhost:11111}. This server presents a complete, network-addressable hardware suite mimicking a complete observatory with: a camera, telescope mount, focuser, dome, filter wheel, weather station, and safety monitor. The underlying \texttt{pytest} suite runs Astra against this simulator with zero mocks injected at the device interface. Every outbound Alpaca HTTP request, internal \texttt{multiprocessing.Pipe} message, and SQLite database transaction traverses the identical code execution paths deployed in live production environments.

This mock-less integration strategy comprehensively validates all supported schedule action types, runtime watchdog state transitions, asynchronous weather-alert injections mid-sequence, multi-cadence FITS header polling, and the internal database queue. For action-level validation, the suite programmatically instantiates a live \texttt{Observatory} subclass, writes a single-line JSONL schedule payload to disk, and tracks the file-watchdog's event detection loop before asserting on the final state machine and absence of triggered \texttt{report\_device\_issue()} routines.

\section{DEPLOYMENT AND OPERATIONS}
\label{sec:deployment}

Astra reached production in January 2024. The SPECULOOS-South rollout across the four 1\,m telescopes at ESO Paranal (Chile) proceeded in phases timed to coincide with the annual service mission: Callisto in January 2024, Ganymede in March 2025, and Europa and Io in February 2026. Each telescope runs its own Astra process in fully autonomous robotic mode. SAINT-EX, a 1\,m telescope at the Observatorio Astron\'{o}mico Nacional at San Pedro M\'{a}rtir (Mexico), has operated Astra since January 2024 for its transit survey programme. The ETH Observatory, a 0.5\,m facility in Zurich (Switzerland), has operated Astra since January 2024 for autonomous student observing programmes.

At SPECULOOS-South, each Astra instance runs on a dedicated Windows Server 2022 host (Lenovo ThinkSystem SR250~V2, Intel Xeon E-2378, 16\,GB ECC RAM, 2\,TB in RAID~10). 
The choice of Windows was dictated by the site's existing instrumentation: several devices expose ASCOM COM drivers with no native Alpaca equivalent. The ASCOM Remote\footnote{\url{https://github.com/ASCOMInitiative/ASCOMRemote}} server, provided by the ASCOM Initiative, translates these COM calls into the HTTP/REST Alpaca interface that Astra expects, and this translation layer has proved reliable in production. 
Monitoring over a five day period, the main process had a mean CPU usage of 36\%, with peaks of 293\% during periods of high activity such as image acquisition, plate solving, or header completion. 
Each device subprocess had a mean CPU usage of under 1\%, with peaks of 417\% during active operations. 
The combined CPU usage across all processes has a mean of 38\% and reaches 833\% at its peak. Memory usage is stable: the main process occupies about 353\,MB and each device subprocess 71--74\,MB, for a total of approximately 857\,MB across all processes.

Astra has maintained continuous autonomous operation at each site since commissioning, with no schedule aborts attributable to Astra software failures recorded at any facility. All interruptions to scheduled observations trace to hardware faults outside Astra's control, principally instrument failures and site infrastructure issues.

\section{ON-SKY PERFORMANCE}
\label{sec:performance}

The measurements reported here are drawn from the SPECULOOS-South network, which provides the longest operational baseline and the largest science-frame archive of the Astra deployments.

\subsection{Autoguiding Performance}
\label{sec:perf-guiding}

Guiding performance was assessed across all four SPECULOOS-South telescopes using science sequences acquired during January and May 2026. For each sequence, the sigma-clipped standard deviation of the pointing displacement was computed from the per-frame \texttt{donuts} shift estimates stored in the guiding log. The median sigma-clipped displacement standard deviation per telescope was 0.32\,pixels (0.10\unit{\arcsecond}) for Callisto, 0.32\,pixels (0.10\unit{\arcsecond}) for Ganymede, 0.38\,pixels (0.12\unit{\arcsecond}) for Io, and 0.52\,pixels (0.16\unit{\arcsecond}) for Europa. The network median is 0.35\,pixels (0.11\unit{\arcsecond}), well below the 1\,pixel photometric-stability requirement of the SPECULOOS survey.

\begin{figure}[ht]
\centering
\includegraphics[width=\columnwidth]{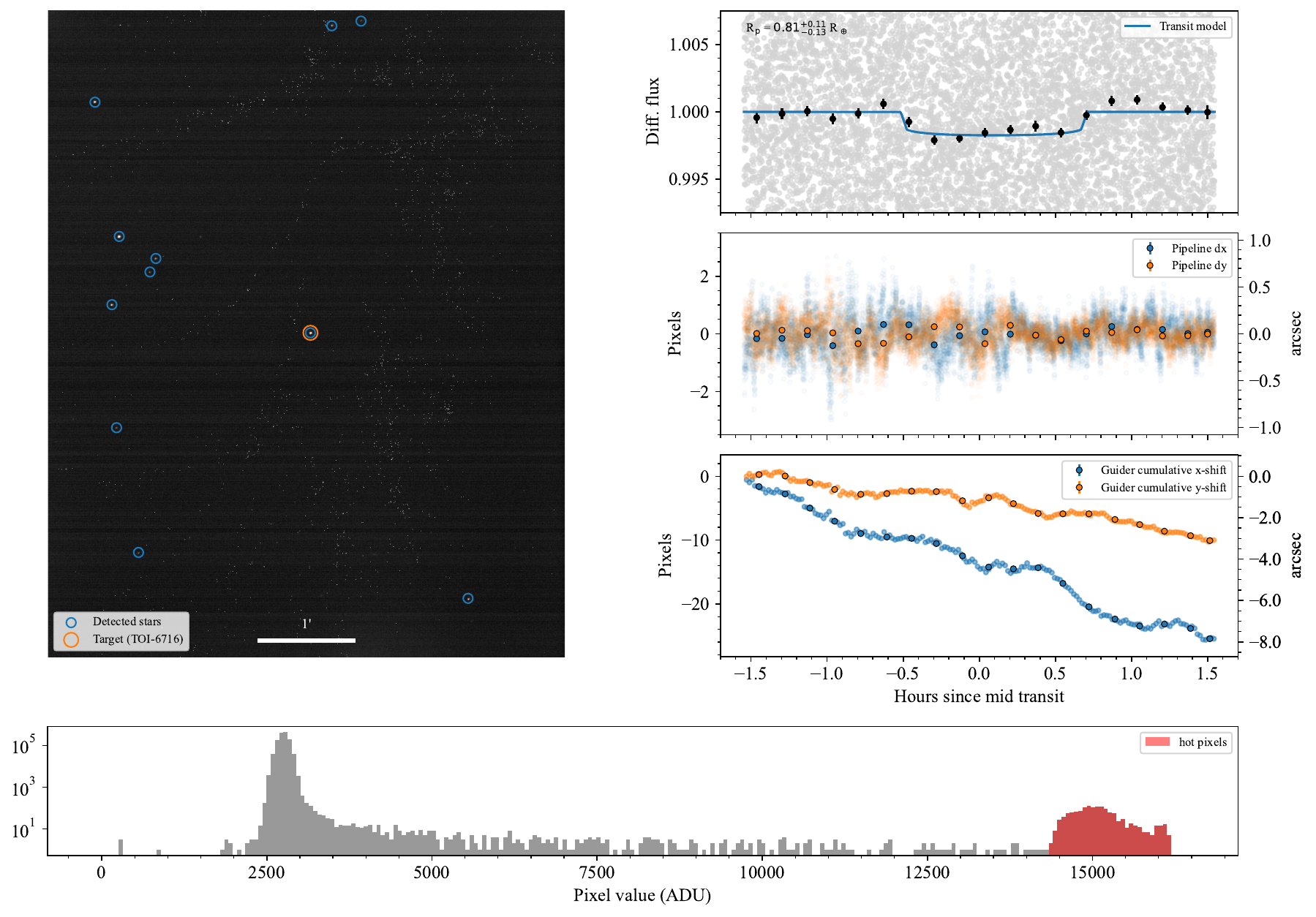}
\caption{Autoguiding performance and detrended transit light curve of TOI-6716.01 observed by SPECULOOS-South Callisto on 11~January~2026 in the $zYJ$ band at 1\,s cadence (10\,121 exposures). \emph{Left}: field image with detected stars (blue circles) and the target TOI-6716 (orange circle); the white bar marks 1\unit{\arcminute}. \emph{Right, top}: differential flux with the best-fit transit model overlaid (fit $R_{\text{p}}$ = $0.81^{+0.11}_{-0.13}$ $R_\oplus$); grey points are individual 1\,s exposures and black points are 7.2\,min bins. \emph{Right, middle}: pipeline-measured centroid displacement in x and y (dx, dy). \emph{Right, bottom}: cumulative guider shift corrections in both axes from \texttt{donuts}. The sigma-clipped two-dimensional pointing standard deviation for this sequence is 0.32\,pixels (0.10\unit{\arcsecond}).}
\label{fig:toi6716-lc}
\end{figure}

A representative single-sequence observation demonstrates the stability achieved under this guiding logic. On 11~January~2026, the Callisto telescope observed the transit of TOI-6716.01~\cite{scott2026two} (total transit duration of 1.237$\pm$0.015\,h) in the $zYJ$ band, acquiring 10\,121 consecutive 1\,s exposures. Despite hot pixels occupying approximately 0.3\% of the detector~\cite{pedersen2024spirit}, the donuts-based guider maintained a stable pointing standard deviation of 0.32\,pixels (0.10\unit{\arcsecond}) throughout the sequence. Figure~\ref{fig:toi6716-lc} shows the pipeline-measured centroid residuals and the corresponding guider corrections alongside the detrended differential light curve.

\subsection{Plate Solving Performance}
\label{sec:perf-platesolve}

Plate-solve success was assessed from the full operational logs of all four SPECULOOS-South telescopes. For Europa, Io, and Ganymede, the solve failure rate was below 1\% across all scheduled pointing corrections. Failures were caused by poor observing conditions (cloud obscuration), camera shutter faults, and satellite transits in the acquisition frame. Callisto carries a camera with a narrower field of view and a higher noise floor than the other three telescopes; the smaller number of detected stars per frame raises its failure rate to 3\%. Across all successful solves, the median residual pointing offset after correction was within 0.01' of the nominal target coordinates. This residual is likely limited by the mount's repointing precision, repointing is not attempted if the target is within the user defined tolerance (default 0.1').

\section{CONCLUSIONS}
\label{sec:conclusions}

Astra demonstrates that an open-source, standards-based software stack can sustain fully autonomous survey operations at production scale. The system controls all observatory hardware through the ASCOM Alpaca protocol, isolates each device in its own subprocess, and coordinates scheduling, safety supervision, plate solving, autofocus, and autoguiding from a single Python process without external message-broker infrastructure. A FastAPI web service exposes the same telemetry through a browser interface and through REST and WebSocket APIs, so operators and external monitoring systems read identical state. Since January 2024 Astra has run unattended on six telescopes across the SPECULOOS-South network, SAINT-EX, and the ETH Observatory, and no facility has recorded a schedule abort attributable to Astra. Across SPECULOOS-South the system holds a median pointing scatter of 0.11\unit{\arcsecond} and keeps plate-solve failure rates below 1\% on three of the four telescopes. These results replace a Windows-locked proprietary toolchain with a cross-platform stack that the SPECULOOS consortium now maintains in the open, and is publicly available at: \url{https://github.com/ppp-one/astra}.

\acknowledgments 
 
The ULiege's contribution to SPECULOOS has received funding from the European Research Council under the European Union's Seventh Framework Programme (FP/2007-2013) (grant Agreement n$^\circ$ 336480/SPECULOOS), from the Balzan Prize and Francqui Foundations, from the Belgian Scientific Research Foundation (F.R.S.-FNRS; grant n$^\circ$ T.0109.20), from the University of Liege, and from the ARC grant for Concerted Research Actions financed by the Wallonia-Brussels Federation. MG is F.R.S-FNRS Research Director. 
This research is supported by the Science and Technology Facilities Council (STFC; grant n$^\circ$ ST/S00193X/1, ST/W002582/1, and ST/Y001710/1).

\bibliography{report} 
\bibliographystyle{spiebib} 

\end{document}